\documentclass[pra,twocolumn,showpacs,amsmath,amssymb,superscriptaddress,floatfix]{revtex4-1}
\usepackage{graphicx,color}
\usepackage{amsmath,amssymb,bm}
\usepackage{braket}		
\usepackage{ulem}
\usepackage{epstopdf}
\definecolor{darkred}{rgb}{0.90,0,0}
\definecolor{darkgreen}{rgb}{0,0.60,.2}
\definecolor{darkblue}{rgb}{0,0,1}
\definecolor{grey}{cmyk}{0,0,0,0.25}
\definecolor{orange}{cmyk}{0,0.6,0.8,0}

\newcommand{\be}{\begin{equation}}
\newcommand{\ee}{\end{equation}}

\newcommand{\avg}[1]{\overline{#1}}
\newcommand{\rmi}{\mathrm{i}}
\newcommand{\rmd}{\mathrm{d}}
\newcommand{\rme}{\mathrm{e}}

\begin{document}
\title{
Phase diagram of the 3D Anderson model for uncorrelated speckle potentials}

\author{M. Pasek}
\affiliation{Laboratoire Mat\' eriaux et Ph\' enom\`enes Quantiques,
Universit\' e Paris Diderot-Paris 7 and CNRS, UMR 7162, 75205 Paris Cedex 13, France}
\affiliation{Laboratoire Kastler Brossel, UPMC-Sorbonne Universit\'es, CNRS, ENS-PSL Research University, Coll\`{e}ge de France, 4 Place Jussieu, 75005 Paris, France}

\author{Z. Zhao}
\affiliation{Laboratoire Mat\' eriaux et Ph\' enom\`enes Quantiques,
Universit\' e Paris Diderot-Paris 7 and CNRS, UMR 7162, 75205 Paris Cedex 13, France}

\author{D. Delande}
\affiliation{Laboratoire Kastler Brossel, UPMC-Sorbonne Universit\'es, CNRS, ENS-PSL Research University, Coll\`{e}ge de France, 4 Place Jussieu, 75005 Paris, France}

\author{G. Orso}
\affiliation{Laboratoire Mat\' eriaux et Ph\' enom\`enes Quantiques,
Universit\' e Paris Diderot-Paris 7 and CNRS, UMR 7162, 75205 Paris Cedex 13, France}

\begin{abstract}
We investigate the localization properties of atoms moving in a three-dimensional optical lattice in the presence of an uncorrelated disorder potential 
having the same probability distribution $P(V)$ as laser speckles.
We find that the disorder-averaged (single-particle)  Green's function, calculated via the coherent potential approximation, is in very good agreement with 
exact numerics.  
Using the transfer matrix method, we compute the phase diagram
in the energy-disorder plane and show that its peculiar shape  can be  understood  from the  self-consistent theory of localization. 
In particular, we recover the large asymmetry in the position of the mobility edge for blue and red speckles, which was recently observed numerically for correlated speckle potentials.
\end{abstract}

\maketitle

 \section{Introduction and model Hamiltonian }
\label{sec:intro}

Recently there has been a growth of experimental and theoretical studies \cite{SanchezPalencia:DisorderQGases:NP10} on Anderson localization of ultra-cold atoms
exposed to disordered optical potentials, including quasi-periodic lattices and blue-detuned laser speckles.
The latter behave quite differently from typical models of randomness 
considered for condensed matter systems. Indeed, the potential distribution $P(V)$ of blue speckles is not Gaussian but follows the exponential (Rayleigh) law \cite{Goodman:07,Shapiro:JPA:2012}:
\begin{equation}
 P(V) = \frac{\Theta(V+V_0)}{V_0}\ \exp{\left(-\frac{V+V_0}{V_0}\right)}
\label{Eq:Rayleigh-up}
\end{equation}
where  $\Theta$ is the Heaviside function and $V_0$ is the disorder strength. In Eq.~(\ref{Eq:Rayleigh-up}), we have shifted the potential by its average 
value so that $\avg{V}\!=\!0$. 
This distribution is bounded from below by $-V_0$ and is asymmetric, implying that odd moments of the potential will be non zero. The probability distribution of red-detuned potentials, which have not yet been implemented experimentally, is obtained by simply changing $V$ to $-V$ in Eq.~(\ref{Eq:Rayleigh-up}).
A second fundamental aspect of optical speckles is that they are spatially correlated, with a typical grain size 
of the order of 1 $\mu$m.

Three
different experiments \cite{Jendrzejewski:AndersonLoc3D:NP12,Semeghini:3DAnderson:NP15,Kondov:ThreeDimensionalAnderson:S11} claimed the observation of 3D Anderson localization of atoms exposed to blue speckles. However, the estimated
mobility edge -- namely the critical value $E=E_c$ of the energy separating localized ($E<E_c$) from extended ($E>E_c$) states -- is in all cases larger (and even much larger in \cite{Kondov:ThreeDimensionalAnderson:S11}) than the 
current theoretical and numerical predictions~\cite{Delande:MobEdgeSpeckle:PRL2014,Pilati:LevelStats:2015,Kuhn:Speckle:NJP07,Piraud:MobilityEdge3D:NJP13,Yedjour:MobilityEdge3D:EPJD10}. 
The question arises  whether the problem is due to inadequate experimental measurements (it is for example very difficult 
to correctly analyze the density profiles after a relatively
short time expansion in the presence of disorder~\cite{Mueller:CommentDeMarco:PRL14}) or to incorrect
theoretical predictions for the mobility edge.

From the theory side, exact  numerical calculations of the mobility edge obtained via the transfer-matrix 
technique \cite{Delande:MobEdgeSpeckle:PRL2014} have revealed a discrepancy with
 previous estimates based  on the self-consistent theory of Anderson localization (SCTL) \cite{Kuhn:Speckle:NJP07,Yedjour:MobilityEdge3D:EPJD10,Piraud:MobilityEdge3D:NJP13}. In particular it was found that 
the on-site potential distribution
is extremely important. Indeed there is a large difference between blue-detuned speckles, where the mobility edge
is systematically negative (i.e.\ below the average potential), and red-detuned speckles, where
it changes from negative at low $V_0$ to positive at large $V_0$ \cite{Delande:MobEdgeSpeckle:PRL2014}.
On the other hand, the details of the spatial correlation function -- beyond the correlation length
which fixes the characteristic quantum energy scale -- do not play a major role.
The aforementioned discrepancy with SCTL predictions originates most probably from the lack of accuracy in the calculation of the 
disorder-averaged single-particle Green's function, which is a fundamental ingredient of the SCTL approach. Indeed, in Refs.~\onlinecite{Kuhn:Speckle:NJP07,Yedjour:MobilityEdge3D:EPJD10,Piraud:MobilityEdge3D:NJP13}, this quantity 
   has been computed within the self-consistent Born approximation (SCBA), which 
 is by construction insensitive to the color of the speckle, and can only apply for a sufficiently weak disorder.  
  
 In this work we disentangle the role of the on-site potential distribution from the effects of spatial
correlations by studying a spatially uncorrelated 3D Anderson model with a Rayleigh potential distribution.
We calculate the Green's function numerically, and show that the obtained results are well reproduced by the coherent-potential approximation (CPA), which takes into account the full statistical properties of the potential. 
The CPA self-energy
 is then injected in the SCTL to estimate the position of the mobility edge, which turns out to be in reasonably good
agreement with the exact transfer-matrix calculations. 
It is worth mentioning that the CPA method has not yet been generalized to spatially correlated speckle potentials, although this generalization has been attempted for other types of spatially-correlated disorder \cite{Jarrell:RMP2005,Zimmermann:ExtendedCPA:PRB2009}.
  
In first quantization, the 3D Anderson model is given by 
\begin{equation}\label{Ham}
H=  \sum_{\langle ij\rangle} -J |i\rangle \langle j|+\sum_i V_i |i\rangle \langle i |,
\end{equation}
where indices $i, j$ label the sites of the lattice, $J$ is the hopping term between nearest neighbors, and $V_i$ the random external potential which is assumed to be uncorrelated, $\avg{V_i V_j}=\langle V^2\rangle \delta_{ij}$, and obey the potential distribution in Eq.~(\ref{Eq:Rayleigh-up}).
For convenience, in the following we will use units $J=1$ and  $\hbar=1$.
For the specific case of the Anderson model, there is no need to study separately the blue and red-detuned speckles.
Indeed,
the cubic lattice with nearest-neighbor coupling being bipartite, the sign of $J$ is irrelevant and does not affect
localization properties. As a consequence, a change $V\to -V$ is equivalent to reversing the sign of the Hamiltonian, $H\to -H$.
This means that all results obtained here for the blue speckle with the Rayleigh distribution, including the phase diagram, apply also to the red speckle under the change $E\to -E$, as shown in Fig.~\ref{fig:tm}. 
In other words, by studying the two mobility edges around $E=-6$ and $E=+6$,
one covers both the blue and red-detuned cases respectively.

The plan of the paper is as follows. In Sec.~\ref{sec:TM} we present the exact phase diagram for the 3D Anderson model 
with the Rayleigh potential distribution, Eq.~(\ref{Eq:Rayleigh-up}), obtained numerically via the transfer-matrix technique.
In Sec.~\ref{sec:GF} we calculate the disorder-averaged single-particle Green's function numerically, and analytically using the CPA and the SCBA.  
The obtained results are then used in Sec.~\ref{sec:sctAL}  to estimate the position of the mobility edge via the 
SCTL and to compare it with the exact numerical results of  Sec.~\ref{sec:TM}.
Finally, we examine in Sec.~\ref{sec:cold_atoms} how our findings help to interpret numerical and experimental results
for cold atoms in a continuous speckle potential.

\section{Exact phase diagram}

\label{sec:TM}
\begin{figure}
\includegraphics[width=0.94\columnwidth]{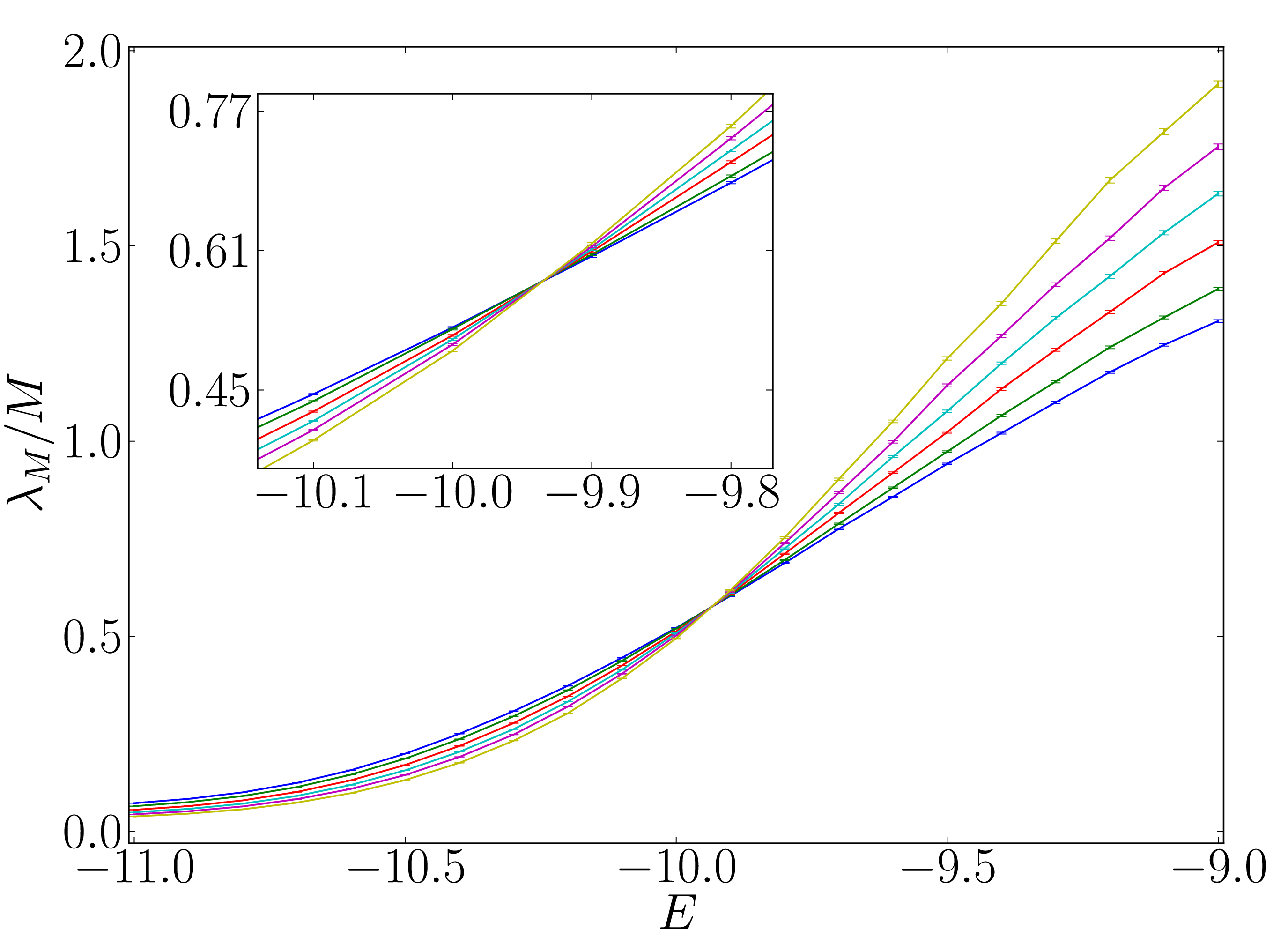}
\caption{(Color online) Transfer-matrix results for the 3D Anderson model with  uncorrelated, blue-detuned
speckle potential of  strength $V_0=8.$ Each curve displays the ratio of the localization length $\lambda_M$ of a long bar
with cross-section $M\times M$ to the bar transverse size $M$, as a function of energy.
The various curves from $M=16$ (least steep curve) to $M=31$ (steepest curve), cross at the mobility edge $E_c \approx -9.93$.
}\label{fig:crossingtm}
\end{figure}

\begin{figure}
\includegraphics[width=.9\columnwidth]{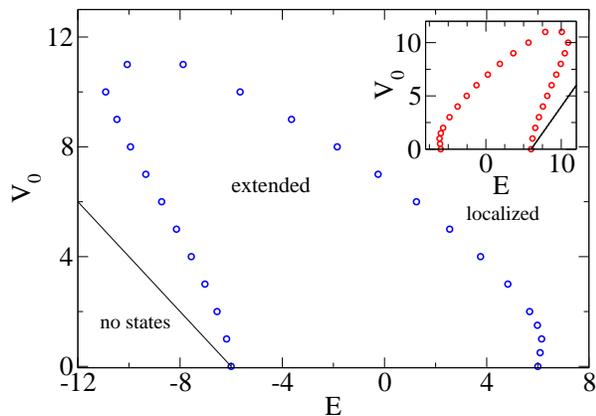}
\caption{(Color online) Exact localization phase diagram for the 3D Anderson model with blue-detuned speckle potential obeying the
Rayleigh distribution, Eq.~(\ref{Eq:Rayleigh-up}), as obtained from the transfer-matrix technique and finite-size scaling. 
Notice that  there are rigorously  no states below the solid line corresponding to $E=-6-V_0$.
The phase diagram for red-detuned speckles is simply obtained under the change $E \to -E$, as shown in the 
 inset.}
 \label{fig:tm}
\end{figure}

The transfer-matrix method consists in computing recursively the conductance of a bar with transverse section $M\times M$ sites, and length $L \gg M$. The logarithm of the conductance (averaged over disorder realizations) 
decays linearly with the length of the bar, giving access to the quasi-1D localization length $\lambda_M$.
The scaling theory of localization indicates that $\lambda_M/M$ decreases with $M$ in the localized regime,
increases in the diffusive regime, and is constant for large $M$ at the mobility edge \cite{McKinnonKramer:TransferMatrix:ZPB83}. The various $\lambda_M/M$ vs.~energy curves -- for various $M$ values -- thus cross at the critical energy, the mobility edge, as shown in Fig.~\ref{fig:crossingtm}.
By gathering  results at various values of the disorder strength $V_0,$ we  obtain
the phase diagram of the Anderson model, shown in Fig.~\ref{fig:tm} for a blue-detuned speckle.

We note that the trajectory of the mobility edge 
behaves very differently at the two band edges. Indeed, starting from the left band edge of the clean model, $E=-6$, the mobility edge shifts to lower and lower energy as $V_0$ 
increases until $V_0\approx 10$, which is  close to the critical disorder  needed to localize all states. 
Since  the potential distribution has a sharp cut-off at $V=-V_0$, there are strictly no states
below $E=-6-V_0$. In contrast, the spectrum is not bounded from above.
At the right band edge, $E=+6$, however, the mobility edge moves only slightly to the 
right for very weak disorder, reaching $E=6.15$ at $V_0=1$, 
 and then moves backward in a monotonous way. 
This means that 
even for moderate disorder, starting from $V_0\approx1$, the asymmetry of the Rayleigh probability distribution $P(V)$
becomes important and must be taken into account. This rules out the use of the SCBA for the calculation of the 
 disorder-averaged Green's function, as we will show below.

\section{Disorder-averaged Green's function}
\label{sec:GF}
In this section we calculate the disorder-averaged single-particle Green's function  numerically, and analytically 
using two different approximation schemes, namely the SCBA and the CPA. We also discuss two related quantities,  
the (disorder-averaged) density of states and the effective band edge (neglecting Lifshitz tails).
 
In the absence of disorder $(V_0=0)$, the Hamiltonian $H$ becomes diagonal in momentum space: $H=H_0=\sum_\mathbf k \epsilon(\mathbf k) |\mathbf k\rangle \langle \mathbf k|$, 
where $\epsilon(\mathbf k)=-2(\cos k_x+\cos k_y+\cos k_z)$ is the energy dispersion of the atom in the cubic lattice.
Hence the diagonal part of the free-particle Green's function in configuration space representation $G_0(E)\equiv \langle n|(E-H_0)^{-1}| n\rangle$  is translationally invariant and 
given by 
\be\label{Green0}
G_0(E)=\int_{-\pi}^\pi \frac{\rmd^3k}{(2\pi)^3} \frac{1}{E-\epsilon(\mathbf k)+\rmi 0}.
\ee
 Following Joyce \cite{Joyce:GreenElliptic:1998},
we can express the unperturbed Green's function of a cubic lattice as  $G_0(E)=P(6/E)/E$, where 
\be\label{Pz}
P(z)=\frac{1-9\xi^4}{(1-\xi)^3(1+3 \xi)}\left [\frac{2}{\pi} K(k_1)\right]^2.
\ee
Here $\xi$ and $k_1$ are functions of $z$ defined as  
\begin{eqnarray}
\xi(z)&=&\left (\frac{1-\sqrt{1-z^2/9}}{1+\sqrt{1-z^2}}\right)^{1/2},\\
k_1(z)^2&=&\frac{16 \xi^3}{(1-\xi)^3(1+3\xi)}, 
\end{eqnarray}
and $K$ is the complete elliptic integral of the first kind.

\subsection{Numerical computation and analytical estimates of the self-energy}
In the presence of a random potential,  the translational symmetry is restored only after averaging over different realizations.
The corresponding disorder-averaged Green's function is given by
\be\label{Green}
\avg{G(E)}=\int_{-\pi}^\pi \frac{\rmd^3k}{(2\pi)^3} \frac{1}{E-\epsilon(\mathbf k)-\Sigma(E,\mathbf k)},
\ee
where $\Sigma=\Sigma^\prime+\rmi\Sigma^{\prime\prime}$ is the self-energy.
The real part of the self-energy $\Sigma^{\prime}(E,\mathbf k)$ represents how the energy of a plane wave with wavevector $\mathbf{k}$ is shifted from $\epsilon(\mathbf k)$ (``renormalized'') under the
influence of the disorder, while the imaginary part $\Sigma^{\prime\prime}(E,\mathbf k)$ -- which is always negative -- yields the broadening of the energy distribution \cite{Mahan:Many-Particle}.

\begin{figure}
\includegraphics[width=.96\columnwidth]{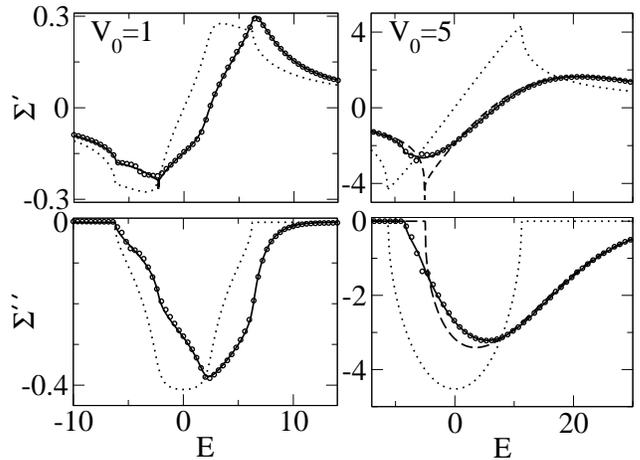}
\caption{ Energy dependence of the real $(\Sigma^{\prime})$ and imaginary $(\Sigma^{\prime\prime})$ parts of the self-energy calculated using the coherent potential approximation (solid line, Eq.~(\ref{CPA})) and the self-consistent Born approximation (dotted line, Eq.~(\ref{SCBA})) for two values of the disorder strength: $V_0=1$ (left) and $V_0=5$ (right).  Also shown is the atomic limit  approximation (dashed line, from Eqs.~(\ref{CPA2}, \ref{CPA3})) for  $V_0=5$. 
In both cases, the CPA is a truly excellent approximation to the exact numerical results (for $k=0$) plotted as circles, reproducing
almost all details of both the real and imaginary parts of the self-energy (note that for the numerical computations, system sizes of $M^3$ with $M=28$ were used). In contrast, the SCBA is a rather poor
approximation, even at moderate disorder strength $V_0=1$ where it is supposed to work best.}
\label{fig:selfenergyversusE}
\end{figure}

Numerically computing the self-energy is possible, but not completely straightforward.
The starting point is the 
temporal representation of the Green's function as:
\begin{equation}
\frac{1}{E-H+\rmi 0} = -\rmi \int_0^{\infty}{\rme^{-\rmi Ht}\ \rme^{\rmi Et}\ \rmd t}. 
\end{equation}
As mentioned previously, the \textsl{average} Green's function in presence of disorder is invariant by translation, i.e.\ diagonal in momentum space. Thus, one has:
\begin{equation}
\begin{split}
\avg{\langle \mathbf k|G(E)|\mathbf k \rangle}& = \frac{1}{E-\epsilon(\mathbf k)-\Sigma(E,\mathbf k)}\\ 
                                              & =  - \rmi \int_0^{\infty}{\avg{\langle \mathbf k|\rme^{-\rmi Ht}|\mathbf k \rangle} \ \rme^{\rmi Et}\ \rmd t}.
\end{split}
\end{equation}
The numerical calculation then amounts to propagating an initial plane wave $|\mathbf k\rangle$ with the disordered Hamiltonian $H$ during time $t$ (with periodic boundary conditions) and to computing
the overlap of the time-evolved state with $\langle \mathbf k|,$ followed by a Fourier transform from time to energy; subtracting  $E-\epsilon(\mathbf k)$ from the inverse of the result yields the self-energy.
This procedure is then repeated for several independent realizations of the disorder  to perform disorder averaging. In order to obtain small statistical fluctuations, a rather large number (of the order of $10^5$) of  disorder realizations is needed. 
Moreover,  a sufficiently large system -- much larger than the mean free path -- has to be used to reduce finite-size effects. 
In our numerical results, we found that the self-energy has a very weak dependence on  momentum, which is expected for spatially uncorrelated potentials, $\avg{V_i V_j}=V_0^2 \delta_{ij}$. 
In Fig.~\ref{fig:selfenergyversusE}, we plot our numerical results  for the real (upper panel) and imaginary (lower panel) parts of the self-energy, calculated at $\mathbf k=0$ for two different values of the disorder strength.

In the following we compare our exact numerics with several approximate methods to calculate the self-energy. By construction, these methods yield estimates which are independent of momentum, $\Sigma(E,\mathbf k)=\Sigma(E)$.
The simplest one is the Born approximation, which is given by:
\be\label{BA}
\Sigma(E)=V_0^2 G_0(E).
\ee
A slight, but simple, improvement is the self-consistent Born approximation (SCBA) where the disorder-free Green's function in Eq.~(\ref{BA}) is modified self-consistently, leading to the equation:
\be\label{SCBA}
\Sigma(E)=V_0^2 G_0(E-\Sigma).
\ee
This quantity can be easily calculated numerically by successive iterations, starting from the
Born approximation, Eq.~(\ref{BA}).
Since only the second moment of the potential distribution appears in the rhs of Eq.~(\ref{SCBA}), the SCBA self-energy \emph{does not} depend on the details of the potential distribution $P(V)$, which makes sense only for  sufficiently weak disorder. 

A better approximation scheme at stronger disorder is given by the CPA \cite{Elliott:CPAreview:RMP1974}. 
 The basic idea of the CPA is as follows (for a pedagogical discussion see Ref.~\onlinecite{Rickayzen:Greenbook}): one isolates a single site $i$ where the potential is chosen randomly
according to $P(V)$, and replaces the surrounding sites by an effective homogeneous medium characterized by a uniform
self-energy. One can then compute the single-site $t$-matrix of site $i$ embedded in the surrounding effective medium.
The consistency condition is that the $t$-matrix, averaged over the potential distribution at site $i,$ vanishes; this translates
in the self-consistent equation
\be\label{CPA}
\int P(V) \frac{V-\Sigma(E)}{1-(V-\Sigma(E))G_0(E-\Sigma(E))}\ \rmd V\ =\ 0,
\ee
which clearly depends on the full probability distribution of the potential.
For weak disorder, the CPA reduces to the SCBA, as is clear  from Eq.~(\ref{CPA}) using the Taylor 
expansion $1/(1-x)\simeq 1+x$, with $x=(V-\Sigma)G_0(E-\Sigma)$. As the disorder strength increases, however, higher order terms in the expansion become
important and  all moments of the potential distribution (cubic, quartic, etc.) will start to play a role. Hence, in general, the CPA yields more accurate results than the SCBA, as previously known for the Anderson model with random-box potential \cite{Kroha:SelfConsistentTheoryAnderson:PRB90,Kroha:SCTL:Physica90}. 
By substituting the rhs of  Eq.~(\ref{Eq:Rayleigh-up}) in Eq.~(\ref{CPA}) and performing the integration over $V$, we obtain 
\be\label{CPAanalytics}
e^{-u}(E_i(u)-\rmi \pi)=g V_0,
\ee
where $g=G_0(E-\Sigma)$, $u=(1+g(V_0+\Sigma))/(gV_0)$ and $E_i(u)$ is the exponential integral function defined as  $E_i(u)=-\textrm{PV}\int_{-u}^\infty \rmd t e^{-t}/t$, $\textrm{PV}$ being the principal value.
Equation (\ref{CPAanalytics}) can be easily  solved numerically by a root-searching algorithm for the complex variable $\Sigma$ or, again, by an iteration scheme.

 The SCBA and CPA results for the self-energy are shown in Fig.~\ref{fig:selfenergyversusE}  as
dotted and  solid lines, respectively.
The most striking observation is
that the CPA gives excellent predictions, almost in perfect agreement with our numerical results. Hence, this 
approximation  can be safely used in the SCTL to compute the phase diagram of the Anderson model with a potential distribution typical of optical speckles.
In contrast, the SCBA is a rather poor approximation, even at moderate disorder strength.

By construction, in the SCBA the real (imaginary) part of the self-energy is an odd (even) function of the energy and of the potential strength. We see in
Fig.~\ref{fig:selfenergyversusE} that this feature 
disappears  in the CPA,  due to the asymmetry of the Rayleigh distribution, Eq.~(\ref{Eq:Rayleigh-up}). 
In particular, for a fixed disorder strength, the imaginary  part of the  self-energy is much larger (in modulus) around $E=6$ as compared to $E=-6$, suggesting that disorder scattering is much stronger 
in red speckles.

The dependence of the self-energy on the speckle color is most evident in the atomic limit, either for $V_0\gg 1$ or $ E\gg 1$, where all states are deeply localized. By neglecting 
the tunneling term in the Hamiltonian, one is left with a single-site problem. The disorder-averaged Green's function then takes the simple form
\be\label{CPAstrong}
\avg{G(E)}=\int P(V)\ \frac{1}{E-V+i 0} \ \rmd V.
\ee
By substituting the   rhs of  Eq.~(\ref{Eq:Rayleigh-up}) in Eq.~(\ref{CPAstrong}) and performing the integration over $V$, we find 
\be\label{CPA2}
\avg{G(E)}=\frac{1}{eV_0}\left[f\left(\frac{E}{V_0}\right)-\rmi \pi e^{-E/V_0}\Theta(E+V_0)\right],
\ee
where 
\be\label{CPA3}
f(x)=\mathrm{PV}\int_{-1}^{+\infty}\frac{e^{-z}}{x-z} \rmd z .
\ee
The self-energy, calculated from Eqs.~(\ref{CPA2},~\ref{CPA3}),  is shown in Fig.~\ref{fig:selfenergyversusE} 
 for $V_0=5$ as a dashed line and we see that it agrees with the exact numerics at high energies.

We stress that the exact energy spectrum is bounded from below by $E=-6-V_0$, which implies that
the imaginary part of the self-energy ($\Sigma^{\prime\prime}$) is strictly zero below this value. However, numerical results show that there are very few states
immediately above $E=-6-V_0$, in the so-called Lifshitz tail, where  $\Sigma^{\prime\prime}$ 
is extremely small. It only raises at a significantly larger energy $E\approx-8.4$ for $V_0=5$ (while $-6-V_0=-11$), 
see Fig.~\ref{fig:selfenergyversusE} or Fig.~\ref{fig:dos} below. Thus, there is  an effective  band edge which is
higher than the rigorous band edge. 
Both the SCBA and the CPA, being insensitive to Lifshitz tails, have a band edge below which $\Sigma^{\prime\prime}$
vanishes. We see in Fig.~\ref{fig:selfenergyversusE} that the CPA band edge is in excellent agreement with exact numerics, while
the SCBA band edge is displaced towards negative energies.

\subsection{Density of states and band edge}

\begin{figure}
\includegraphics[width=.96\columnwidth]{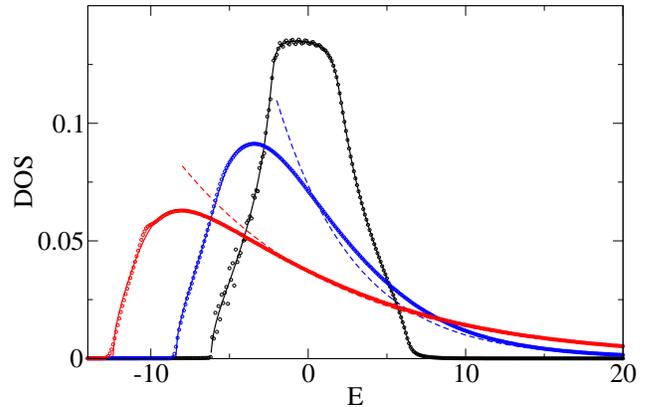}
\caption{(Color online) Density of states calculated in the CPA as a function of energy for three different values of the disorder strength 
$V_0=1$  (black solid line, first from top at $E=0$), $V_0=5$ (blue) and $V_0=10$ (red). The dashed line correspond to the strong disorder
expression in Eq.~(\ref{rhostrong}). Dots correspond to the exact numerical results for system sizes $M^3$ with $M=28$ (for $V_0=1$), and $M=16$ (for $V_0={5,10}$). The CPA is an excellent approximation, reproducing very well the existence of an effective band edge at low energy. 
Only some tiny features above the band edge are not correctly reproduced.
}\label{fig:dos}
\end{figure}

In Fig.~\ref{fig:dos} we plot the averaged density of states (DOS) $\rho=-\textrm{Im}\avg{G(E)}/\pi$ calculated within the CPA for increasing values of the disorder strength, compared to exact numerical values obtained from the self-energy using
Eq.~(\ref{Green}). Again, the agreement between the CPA and  exact numerics is very good.
As disorder increases,  the peak in the DOS shifts towards negative energies and becomes strongly asymmetric~\cite{Trappe:SemiclassicalDOS:JPA2015}. In particular, the DOS
develops exponential tails (decaying as $\exp(-E/V_0)$) at high energies, whereas at negative energies it vanishes more and more sharply with increasing $V_0$. 
From Eq.~(\ref{CPA2}), we indeed see that in the atomic limit
 \be\label{rhostrong}
\rho(E)=\frac{1}{eV_0}e^{-E/V_0}\Theta(E+V_0),
\ee
which coincides with the potential distribution, Eq.~(\ref{Eq:Rayleigh-up}), after replacing $V$ by $E$ . 
As shown in Fig.~\ref{fig:dos}, this  formula correctly reproduces the high-energy tails of the DOS. 
We emphasize that Eq.~(\ref{rhostrong}) can be understood  directly from Eq.~(\ref{CPAstrong}) using the formula $(x+\rmi 0)^{-1}=\mathrm{PV}(1/x)-\rmi\pi \delta (x)$.   
We mention that  the density of states of atoms in spatially-correlated speckle potentials  has been studied both numerically and analytically
in one dimension \cite{Falco:PRA2010}.
  
\begin{figure}
\includegraphics[width=.94\columnwidth]{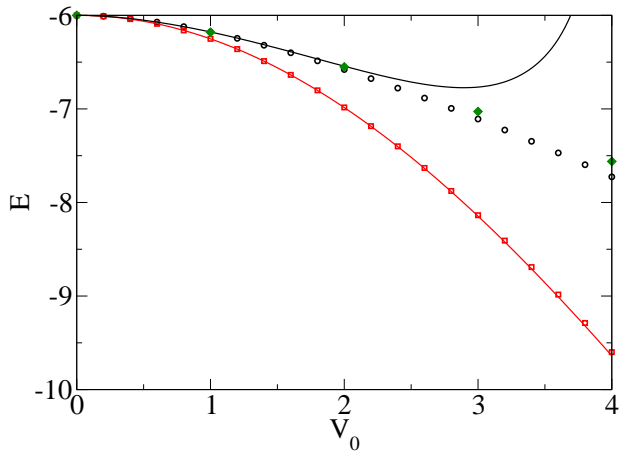}
\caption{(Color online) Band edge $E_{\textrm{BE}}$ plotted as a function of disorder strength, calculated within the SCBA (red squares) and the CPA (black circles). Solid lines refer to
the corresponding approximate analytical expressions, Eqs.~(\ref{EbeSCBA}) and (\ref{bandCPA3}), obtained for weak disorder.  
The exact transfer-matrix results for the mobility edge are also shown (green diamonds).
}
\label{fig:bandedge}
\end{figure}

We now turn our attention to the effective band edge and derive analytical formulas for the SCBA and the CPA predictions 
for weak disorder. 
The unperturbed Green's function is first expanded around the unperturbed band edge $E=-6$, 
yielding $G_0(E)=G_0(-6+\tilde E)\simeq A+B\sqrt{-\tilde E}$, 
where $A=G_0(-6)=(14\sqrt{6}\!+\!20\sqrt{3}\!-\!24\sqrt{2}\!-\!36)K^2[(2\!-\!\sqrt{3})(\sqrt{3}\!-\!\sqrt{2})]/\pi^2\approx -0.252731$ 
and $B=1/(4\pi) \approx 0.079577$. 
Substituting this expression in Eq.~(\ref{SCBA}) gives the self consistent equation:
\be\label{bandSCBA}
\Sigma=V_0^2 \left (A+B\sqrt{\Sigma - \tilde E}\right).
\ee
 For weak disorder, we can replace the self-energy in the rhs of  Eq.~(\ref{bandSCBA}) by its leading term, $\Sigma=V_0^2A$, which is real. 
 As a consequence, the self-energy  can only become complex if the argument of the square root in  Eq.~(\ref{bandSCBA}) becomes negative, 
  that is  
$\tilde E=\Sigma$. This means that the SCBA band edge is given by $E_{\textrm{BE}}^\textrm{SCBA}=-6+AV_0^2+O(V_0^4)$. 
It is possible to get the next order term -- note that by construction of SCBA, only even powers of $V_0$ appear -- using  Eq.~(\ref{bandSCBA}).
By bringing the term proportional to $A$ to the lhs and taking the square of both sides, we find that $\Sigma$ satisfies the following  quadratic equation: 
\be
\Sigma^2-(2AV_0^2+B^2V_0^4)\Sigma+A^2V_0^4+B^2 V_0^4 \tilde E=0.
\ee
The band edge then corresponds to the energy value at which the discriminant vanishes. This gives 
\be\label{EbeSCBA}
E_{\textrm{BE}}^\textrm{SCBA}=-6+AV_0^2+\frac{B^2}{4}V_0^4+O(V_0^6), 
\ee
which is shown in Fig.~\ref{fig:bandedge} with the red solid line.

Let us now focus on the CPA band edge.  
By setting $z=gV_0$, $y=g \Sigma$ 
and $x=V/V_0$, the self-consistent Eq.~(\ref{CPA}) for the CPA self-energy 
can be written as
 \be
 y=\frac{z}{e}\int_{-1}^{+\infty} \frac{e^{-x}x}{1-x z+y} \rmd x.
 \ee
The argument in the integral is expanded in powers of $y$ up to the quadratic term:
\begin{eqnarray}\label{bandCPA1}
y&\simeq &\frac{z}{e}\int_{-1}^{+\infty} e^{-x}x \left(\frac{1}{1-x z}-\frac{y}{(1-x z)^2}+ \frac{y^2}{(1-x z)^3}\right )\rmd x\nonumber \\
  &=& f_0(z)+f_1(z)y+f_2(z)y^2,
\end{eqnarray}  
where $f_i(z)$  are functions of $z$ defined in the interval $[-1,0]$.  For small values of $z$, $f_2(z)\simeq z^2$ implying that $y=F(z)+O(z^6)$, where
\be\label{Effe}
F(z)=\frac{f_0(z)}{1-f_1(z)}.
\ee
By inserting  the asymptotic expansion of the Green's function 
$g \simeq A+B \sqrt{\Sigma-\tilde E}$  in the  formula 
$\Sigma \simeq F(gV_0)/g$ and Taylor-expanding the rhs, we find  with the same level of accuracy that
\begin{equation}
\Sigma\simeq \frac{F(AV_0)+F^\prime(AV_0) BV_0 \sqrt{\Sigma-\tilde E}} {A+B\sqrt{\Sigma-\tilde E}}.
\end{equation}
Repeating the same procedure as above, we obtain the following  approximate 
formula for the CPA band edge: 
\begin{eqnarray}\label{bandCPA3} 
E_{\textrm{BE}}^\textrm{CPA}&=& -6+\frac{F(AV_0)}{A} \\
&+& \left[\frac{F^\prime(AV_0) V_0}{A}  -\frac{F(AV_0)}{A^2}\right]^2 \frac{B^2}{4} +O(V_0^6).\nonumber
\end{eqnarray}
This formula, shown in Fig.~\ref{fig:bandedge} with the black solid line, 
reproduces very well the numerical results (open circles) for the CPA band edge at small disorder.
By using the Taylor expansion $F(z)=z^2+2z^3+7z^4+34z^5+O(z^6),$ one can obtain the expansion
of the CPA band edge in powers of $V_0$:
\begin{eqnarray}\label{bandCPA4}
E_{\textrm{BE}}^\textrm{CPA} & = &
-6 + AV_0^2+2A^2V_0^3+\left(\frac{B^2}{4}+7A^3\right)V_0^4 \nonumber \\
&+&\left(2AB^2+34A^4\right)V_0^5 + O(V_0^6).
\end{eqnarray}
Note that odd-power terms, which are absent in the SCBA, 
appear due to the asymmetry of the Rayleigh potential distribution, and that the fourth order term in Eq.~(\ref{bandCPA4}) is 
different from the SCBA result, Eq.~(\ref{EbeSCBA}).

\section{Self-consistent theory of Anderson localization}
\label{sec:sctAL}

 Starting from the weak-localization corrections to the conductivity (or to the diffusion constant), the SCTL provides 
 a useful microscopic justification of the Ioffe-Regel criterion $k\ell\approx C$ for the onset of localization in 3D continuous models, 
 $C$ being a model-dependent constant of order unity.
 Both the wave number $k$ and mean free path $\ell$ depend on the energy $E$, but not in any simple way. Indeed, the usual relation -- in the absence of disorder -- between $k$ and $E$, $E=\epsilon(\mathbf k)$, is no longer correct
 in the vicinity of the mobility edge (strong-scattering regime), and there is no unique way of defining 
 $k(E)$ and $\ell(E)$. In the simplest approximation, one may assume that the dominant effect of strong disorder
 is to shift the band edge and dispersion relation by the real part of the self-energy $\Sigma^\prime$, so that, for example, $k(E)$ is nothing
 but $k_0(E-\Sigma^\prime(E))$, where the `0' subscript refers to disorder-free quantities. As for the broadening of the
 spectral function (related to the imaginary part $\Sigma^{\prime \prime}$), it is approximately symmetric and thus
 expected to have a negligible effect on averaged quantities.
 For lattice models, which have an anisotropic dispersion relation, $k$ becomes a vector quantity so that the
 Ioffe-Regel criterion cannot be used straightforwardly. A simple generalization has been proposed in Ref.~\onlinecite{Economou:SCTL:1984},
 that reads:
 \be\label{SCTLest}
 S_0(E-\Sigma^\prime)\ell(E)^2= 4\pi C^2,
 \ee
 where $S_0(E)$ is the area of the surface $\epsilon(\mathbf k)=E$ in momentum space:
 \be\label{S0}
 S_0(E)=\int |\nabla _{\mathbf k} \epsilon| \delta(E-\epsilon(\mathbf k)) \rmd^3k.
 \ee
  The mean free path can be written as $\ell(E)\sim v(E)\tau(E)$, where $v(E)=v_0(E-\Sigma^\prime)$ is the average modulus of the particle velocity, defined as
  \be\label{S0v0}
v_0(E)=\frac{1}{S_0(E)} \int |\nabla _{\mathbf k} \epsilon|^2 \delta(E-\epsilon(\mathbf k)) \rmd^3k,
\ee
  and $\tau(E)$ is the relaxation time due to disorder, which is related to the imaginary part of the  self-energy as 
 \be\label{tau}
 \tau=-\frac{1}{2 \Sigma^{\prime \prime}}.
 \ee 
  Then  Eq.~(\ref{SCTLest}) reduces to 
\be\label{Ec}
\frac{S_0(E-\Sigma^\prime)v_0(E-\Sigma^\prime)^2}{\Sigma^{\prime \prime 2} }=16 \pi C^2.
\ee
Note that for continuous systems with a ``massive'' dispersion relation, $E=k^2/2m$, one has $S_0(E)=8\pi mE$ and $v_0(E)=k_0/m=\sqrt{2E/m}$, so that 
Eq.~(\ref{Ec}) indeed reduces to the Ioffe-Regel criterion.

Equations (\ref{S0}, \ref{S0v0}, \ref{Ec}) allow to extract the complete phase diagram of the Anderson model once the CPA self-energy has been computed from Eq.~(\ref{CPA}), and a reasonable choice for the constant $C$ in Eq.~(\ref{Ec}) has been made. For the random-box potential distribution, $P_\textrm{box}(V)=\Theta(W/2-V)/W$, this constant was chosen \cite{Economou:SCTL:1984} to
reproduce the well-known transfer-matrix result $W=W_c=16.5$ for the critical disorder strength at the center of the band ($E=0$). 
Inserting the box distribution in Eq.~(\ref{CPA}) for the CPA self-energy, one obtains $\Sigma_\textrm{box}(E=0)=0-\rmi\,4.7011$ for $W=16.5$, which yields, using Eq.~(\ref{Ec}), the constant $C_\textrm{box}=0.775$.
For definiteness,  we will calculate the mobility edge for the uncorrelated speckle potential using the same constant, that is we set $C=C_\textrm{box}$. Other choices of the constant will produce qualitatively similar results. 

\begin{figure}
\includegraphics[width=.94\columnwidth]{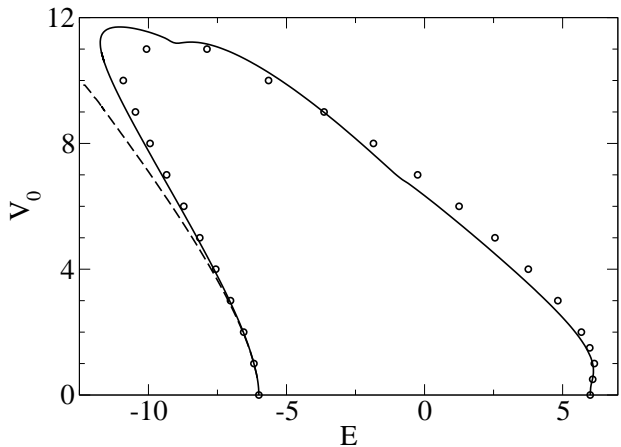}
\caption{ Prediction of the self-consistent theory of localization -- see Eq.~(\ref{Ec}) -- 
for the mobility edge of the Anderson model with Rayleigh potential distribution, using the CPA self-energy (solid line). 
The empty circles correspond to the transfer-matrix results of Fig.~\ref{fig:tm}. 
The dashed line represents the band edge calculated from CPA, neglecting Lifshitz tails.}
\label{fig:sctl}
\end{figure}

In Fig.~\ref{fig:sctl}, we compare the SCTL predictions (solid line) with the transfer-matrix results obtained in Sec.~\ref{sec:TM}. We see that the SCTL reproduces the overall shape of the exact phase diagram. In particular the generalized Ioffe-Regel criterion, Eq.~(\ref{Ec}), 
provides a natural explanation for the behavior of the mobility edge at positive energies. Indeed we see from Fig.~\ref{fig:selfenergyversusE} that near $E=6$
 the imaginary part of the self-energy $\Sigma^{\prime \prime}$ is rather large, or equivalently, the relaxation time in Eq.~(\ref{tau}) is rather short. 
As a consequence, $S_0(E-\Sigma^\prime)$ must therefore increase, which means that $E_c$
will move towards the center of the band, where $S_0(E)$ takes its maximum value, $S_{0}(0)=92.8756$. 

In Fig.~\ref{fig:sctl}, one can also see cusps in the SCTL mobility edge occurring at $E=-9.14$ for $V_0=11.22$
and, although it is less evident, at $E=-0.83$ for $V_0=6.86$.
These features are related to
the presence of Van-Hove singularities \cite{VanHove:Singularity:PR1953} in the DOS of the 
\textsl{clean} system at $E=\pm 2$. Indeed, at the  cusp positions, 
the real part of the self-energy satisfies $E-\Sigma^\prime=-2$ and $E-\Sigma^\prime=2$, respectively. 

Another interesting feature of Fig.~\ref{fig:sctl} is that the mobility edge at negative energies
 remains very close to the CPA band edge, even for moderate disorder. 
This can be easily understood: In the immediate vicinity of the effective band edge $E_{\mathrm{BE}}^{\textrm{CPA}}$, 
the density of states and dispersion relation are very similar to the ones of the disorder-free system near the disorder-free band edge $E=-6,$ meaning that they can be obtained from the disorder-free
quantities by the shift $E\to E-E_{\mathrm{BE}}$, where $E_\mathrm{BE}=E_\mathrm{BE}^\mathrm{CPA}$. This shift is slightly
different from the one in Eq.~(\ref{Ec}), i.e.\ using $\Sigma^\prime$, although the two quantities coincide up to order $V_0^4$ (see above).
Using $E_{\mathrm{BE}}$ is a better approximation however, as it ensures that the density
of states exactly vanishes at $E=E_{\mathrm{BE}}.$ 
Thus, close to the effective band edge, one has approximately the average velocity (the effective mass of the Anderson model is $m=1/2$):
\begin{equation}
v_0(E-E_{\mathrm{BE}})  \simeq 2 k(E),
 \label{Eq:kCPA}
\end{equation}
with $k(E)= \sqrt{E-E_{\mathrm{BE}}}$, whereas the area of the constant-energy shell is
\begin{equation}
 S_0(E-E_{\mathrm{BE}}) \simeq  4 \pi k(E)^2.
 \label{Eq:rhoCPA}
\end{equation}
It also follows that the imaginary part of the self-energy in the Born approximation, Eq.~(\ref{BA}), behaves like:
\begin{equation}\label{eqC}
 \Sigma^{\prime\prime}(E) = - \frac{V_0^2 k(E)}{4\pi}.
\end{equation}
This is in agreement with the square-root behavior of $\Sigma^{\prime\prime}(E)$ above the CPA band edge observed
in Fig.~\ref{fig:selfenergyversusE}. It also means that at low energy the mean free path
is proportional to $1/V_0^2$ and independent of energy. 
Inserting Eqs.~(\ref{Eq:kCPA}--\ref{eqC})  in the generalized Ioffe-Regel criterion, Eq.~(\ref{Ec}), yields
\begin{equation}
 E_{\mathrm{c}}-E_{\mathrm{BE}}=\left(\frac{C V_0^2}{4\pi}\right)^2.
\end{equation}
Therefore, in the weak disorder limit $V_0\ll 1$, the effective band edge is shifted
to the left of $E=-6$ proportionally to $V_0^2$, while the distance to the mobility edge
is much smaller, scaling like $V_0^4$.

\section{Consequences for Anderson localization of cold atoms}
\label{sec:cold_atoms}

A direct implementation of the uncorrelated 3D Anderson model with cold atoms
would require to create a tight cubic optical lattice so that only the
first band is populated, an ultra-cold gas so that $k_B T$ is much smaller than the bandwidth (proportional to the
tunneling rate between neighboring sites), and a speckle potential with a correlation length
much shorter than the lattice spacing. Meeting all these requirements in current experiments seems rather difficult.

From the present study, one can nevertheles
draw a few conclusions about the Anderson localization of massive particles in realistic optical speckle, as realized in recent experiments.
Because of finite-range correlations, the disorder strength $V_0$ must be compared to the other characteristic energy scale, 
namely the ``quantum'' correlation energy $E_{\sigma}=\hbar^2/m\sigma^2$, $\sigma$ being the correlation length.
A thorough discussion of the various possible regimes can be found in Ref.~\onlinecite{Shapiro:JPA:2012}. Especially, the effective band edge 
and the distance of the mobility
edge to the effective band edge were found to scale like $V_0^2$ and $V_0^4$ respectively, in agreement with our results (with however a caveat, see below). 
In the ``quantum'' regime, $V_0\ll E_{\sigma}$, the de Broglie wavelength of the particle is much larger than the correlation
length of the potential, and it is reasonable to expect speckles to essentially behave like $\delta$-correlated potentials.

\begin{figure}[t]
\includegraphics[width=.90\columnwidth]{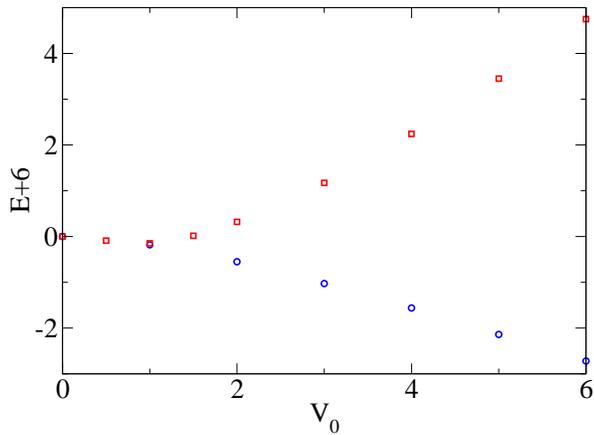}
\caption{(Color online) Zoom-in of the phase diagram (see Fig.~\ref{fig:tm}) for blue (circles) and red (squares) speckles at the bottom of the band. 
The shape is qualitatively similar to the one obtained for spatially correlated speckles 
in Ref.~\onlinecite{Delande:MobEdgeSpeckle:PRL2014}, pointing out the crucial role played by the on-site potential distribution $P(V)$.
 }  
\label{fig:tmzoom}
\end{figure}

In Fig.~\ref{fig:tmzoom} we reproduce the portion of the phase diagram near $E=-6$, calculated in Sec.~\ref{sec:TM} for 
blue and red speckle potentials. Several features of  previous numerical calculations~\cite{Delande:MobEdgeSpeckle:PRL2014} 
for speckle potentials with isotropic correlations are indeed recovered:
\begin{itemize}
 \item At very low $V_0\ll E_{\sigma},$ for both blue and red-detuned speckles, the mobility edge lies below the
 average potential.
 \item For larger $V_0$ and blue-detuned speckle, the mobility edge goes to lower and lower energy.
 \item In contrast, for a red-detuned speckle, the mobility edge has a turning point and becomes larger
 than the average potential energy.
\end{itemize}

However, we stress that a strict mapping of the Anderson model on the behavior of a massive particle is not possible for realistic speckle potentials.
Indeed, as discussed in Ref.~\onlinecite{Shapiro:JPA:2012}, the correlation function of the speckle potential has a long-range tail 
(decreasing not faster than $1/r^2$ at large distance). As a result, the integral of the correlation function   diverges, implying that the ``white noise'' 
limit, which could be associated to a purely uncorrelated potential, does not exist. 
Thus, even if the qualitative behaviors are similar, quantitative differences are expected.

\section{Summary and Conclusion}
\label{sec:sum} 

In conclusion, we have carried out a thorough analysis of the 3D  Anderson model for  uncorrelated random potentials obeying the Rayleigh probability distribution Eq.~(\ref{Eq:Rayleigh-up}). We have shown that the asymmetry in the mobility edge of cold atoms
exposed to blue and red speckles, as recently found numerically in spatially-correlated  isotropic potentials \cite{Delande:MobEdgeSpeckle:PRL2014}, is directly related to the asymmetry of the
Rayleigh distribution under the change $V \to -V$.  
Finally, our work points out the crucial importance of a correct evaluation of the single-particle Green's function in speckle potentials, as a necessary condition to apply the self-consistent theory of localization. We hope that our results will stimulate further work to generalize 
the coherent potential approximation  to spatially correlated speckle potentials. 

This work has been supported by the Region Ile-de-France in the framework of DIM Nano-K.
The authors were granted access to the HPC resources of TGCC under the allocations 2014-057301, 2015-05730 and 2015-057083
made by GENCI (``Grand Equipement National de Calcul Intensif'').

\bibliography{ArtDataBase}

\end{document}